\documentclass[12pt]{amsart}
\usepackage{graphicx,color}
\usepackage{natbib}

\usepackage{amsfonts}

\begin{document}
\bibliographystyle{ametsoc}

\title{Changes in number and intensity of world-wide tropical cyclones}
\maketitle

\begin{center}
{\bf William M. Briggs}  \\ \vskip .05in 
Weill Cornell Medical College \\ 300 E. 71st A3R,
New York, NY 10021 \\ \textit{email:} mattstat@gmail.com \\
\vskip .1in
\today
\end{center}

\newpage

\begin{abstract}
Bayesian statistical models were developed for the number of tropical cyclones and the rate at which these cyclones became hurricanes in the North Atlantic, North and South Indian, and East and West Pacific Oceans.  We find that there is small probability that the number of cyclones has increased in the past thirty years. The {\it rate} at which these storms become hurricanes appears to be constant.  The rate at which hurricanes evolve into category 4 and higher major storms does appear to have increased.  We also investigate storm intensity by measuring the distribution of individual storm lifetime in days, storm track length, and Emanuel's power dissiptation index.  We find little evidence that, overall, the mean of the distribution of individual storm intensity is changing through time, but the {\it variability} of the distribution has increased.  The cold tongue index and the North Atlantic oscillation index were found to be strongly associated with storm quality in the Western, and to a smaller extent, the Eastern Pacific oceans.  The North Atlantic oscillation index was strongly associated with the increase in the rate of strong storms evolving.
\end{abstract}


\section{Introduction}
This paper carries on work we began in \citet{Bri2007}.  In that paper, we statistically characterized the number of tropical cyclones, the chance that these cyclones evolve into  hurricanes, and the intensity of individual storms within a year.  We asked whether or not the number of storms were increasing, whether the chance that they evolved into hurricanes was increasing, and whether the year-by-year mean of intensity was increasing.  We found that it was highly probable that the number of storms had increased, but that the other characteristics reamined constant through the 20th century for the North Atlantic.

Here, we expand this work to include a global statistical model of the North Atlantic, North and South Indian, and West and East Pacific oceans.  Work on this topic is not new, e.g. \cite{WebHol2005}; and e.g. for the Atlantic \cite{Ema2005,Pie2005,Lan2005}.  Much interest naturally centers on whether or not observed increases are due to global warming \cite{Tre2005, TreShe2006, LanHar2006,GolLan2001}.  We do not seek to answer that question here.  We only ask: have the number, and frequency of hurricanes and strong hurricanes that evolve from them, and the distribution of individual storm intensity changed since 1975?  

We use the hurricane reanalysis database (HURDAT) Jarvinen at al. \citeyearpar{JarNeu2002} and, e.g. Landsea et al. \citeyearpar{Lan2004}. The remaining oceans data we got from the ``best track" data on the UNISYS web site (http://weather.unisys.com/hurricane/index.html). This database contains six-hourly maximum sustained 1-minute winds at 10 m, central pressures, and position to the nearest $0.1^\circ$ latitude and longitude from all known tropical storms from 1851-2006.  A cyclone was classified as a ``hurricane" if, at any time during its lifetime, the maximum windspeed ever met or exceeded 65 knots.  Obviously, this cutoff, though historical, is somewhat aribitrary and other numbers can be used: we also use the ``category 4" or ``super typhoon" classification of exceeding 114 knots.  To investigate the realtionship of tropical storms with ENSO, we also use the cold tongue index (CTI) \cite{DesWal1990} and e.g. \cite{Cai2003}.  And we use the North Atlantic oscillation index (NAOI) from \cite{JonJon1997}.

There are obviously observational problems, even since 1975, with basins other than the Atlantic.  Fasullo \citeyearpar{Fas2006} investigated observational quality in the North Atlantic in the context of assessing trends and found that the data was reasonably consistent and that trends could be reliably estimated from it.  Kossin et al. \citeyearpar{KosKna2007} argue the opposite point and say that worldwide the data is rather inconsistent: they propose, and construct, a reanalysis of hurricane activity using satellite observations.  Below, we will see that, particularly with the Indian oceans, the data is observationally biased through time.  We do not seek to correct this bias, and remind our readers that all our results are conditional on the data we use as being correct.  Where we suspect it is not, we give our best guess as to how to interpret the results.

Section 2 lays out the statistical models and methods that we use, Section 3 contains the main results, and Section 4 presents some ideas for future research.

\section{Methods}
\label{secMethods}
We once again adopt Bayesian statistical models.  An important advantage to these models is that we can make direct probability statetments about the results.  We are also able to create more complicated and realistic models and solve them using the same numerical strategy; namely, Gibbs sampling.  We do not go into depth about the particular methods involved in forming or solving these models, as readers are likely familiar with these methods nowadays.  There are also many excellent references available, e.g. \cite{GelCar2003}.

\subsection{Number of storms}

We suppose, in ocean $i$ and year $j$ of $n$, that the number of storms $s_{ij}$ is well approximated by a Poisson distribution as in 
\begin{equation}
s_{ij}|\lambda_{ij} \sim \mbox{Poisson}(\lambda_{ij}). 
\end{equation}
We earlier found that this model well predicted the observed number of storms in the Atlantic \citep[see particularly Figs. 3-4 for a discussion of model fit in][]{Bri2007}. The number of storms in each ocean is assumed independent---this is backed up by the data where any correlation between the $s_{ij}$ for different $i$ is low.   We allow the possibility that $\lambda_{ij}$, the mean number of storms, changes linearly through time, and that the CTI and NAOI may influence, or be associated with the mean CIT \citep{ElsJag2004,ElsBos2001b}.  We use the generalized linear model
\begin{equation}
\label{eq1}
\log(\lambda_{ij}) = \beta_{0i}^s +\beta_{1i}^s t_j + \beta_{2i}^s\mbox{CTI}_j+ \beta_{3i}^s\mbox{NAOI}_j
\end{equation}
The superscipt $s$ is to show we are in the number of storms portion of the model.  The prior for each $\beta_{ki}^s$ is
\begin{equation}
\beta_{ki}^s|\gamma_k, \tau_k \sim \mbox{N}(\gamma_k,\tau_k), \:\:\: k=0,1,2,3
\end{equation}
where $\tau_k$ is the precision (inverse of variance), and to further abstract the mean and variance and thus allow them greater flexability, we use the noninformative priors
\begin{equation}
\gamma_{k} \sim \mbox{N}(0,1e-6),  \:\:\:\:  \tau_{k} \sim \mbox{Gamma}(0.001,0.001).
\end{equation}
Now, it might be that the posterior of $\beta_{1i}^s$  may be mostly or entirely positive for some but not all $i$; this would indicate that the mean number of storms is increasing in those oceans $i$.  It may also be that the posterior of $\beta_{1i'}^s$ may be mostly or entirely negative for other oceans $i'$; this would indicate that the mean number of storms is decreasing in those oceans $i'$.  All $\beta_{1i}^s$, over all oceans, are drawn from the distribution $\mbox{N}(\gamma_1^s,\tau_1)$, and so if the mean of this distribution is greater than 0, then we can conclude that the mean number of storms is increasing across all oceans.  Thus, we can look to the posterior of $\gamma_1^s$ to ascertain whether this is so.

Once a tropical storm develops it, of course, has a chance to grow into a hurricane.  If there are $s_{ij}$ tropical cyclones in ocean $j$ and year $i$ the number of hurricanes is constrained to be between 0 and $s_{ij}$.  Thus, a reasonable model for the number of hurricanes $h_{ij}$ in year $j$ given $s_{ij}$ is 
\begin{equation}
h_{ij}|s_{ij},\theta_{ij} \sim \mbox{Binomial}(s_{ij},\theta_{ij}).
\end{equation}
The parameter $\theta_{ij}$ can be thought of as the proportion of hurricanes that develop from storms.  It is possible, however, as with $\lambda_{ij}$, that $\theta_{ij}$ is dependent on CTI and NAOI and that it changes through time.  To investigate this, we adopt the following logistic regression model
\begin{equation}
\log\left(\frac{\theta_{ij}}{1-\theta_{ij}}\right) = \beta_{0i}^h +\beta_{1i}^h t_j + \beta_{2i}^h\mbox{CTI}_j+ \beta_{3i}^h\mbox{NAOI}_j
\end{equation}
where the superscript $h$ denotes the hurricane portion of the model and we again let the priors and hyperpriors be the same form as in the model for $s_{ij}$.  Like before, we will exmaine the posterior $\beta_{1i}^h$ across each ocean to see whether or not the it is likely that, for ocean $i$, the rate of hurricanes in increasing.  And also if the posterior of $\gamma_1^h$ has most of its probability above 0, we can conclude that the rate of hurricanes is increasing across all oceans.

We carry the model one step further by asking about the chance that a category 4 or above (denoted by $c_{ij}$) hurricane develops from an ordinary hurricane: the number of category 4+ storms is of course constrained to be between 0 and $h_{ij}$.  In analogy with the model above, we have
\begin{equation}
c_{ij}|h_{ij},\xi_{ij} \sim \mbox{Binomial}(h_{ij},\xi_{ij}).
\end{equation}
The parameter $\xi_{ij}$ can be thought of as the proportion of category 4 hurricanes that develop from ordinary hurricanes.   And again, we allow $\xi_{ij}$ to be dependent on CTI, NAOI, and time. Thus,
\begin{equation}
\log\left(\frac{\xi_{ij}}{1-\xi_{ij}}\right) = \beta_{0i}^c +\beta_{1i}^c t_j + \beta_{2i}^c\mbox{CTI}_j+ \beta_{3i}^c\mbox{NAOI}_j
\end{equation}
where the superscript $c$ denotes the category 4+ portion of the model and we again let the priors and hyperpriors be the same form as in the model for $s_{ij}$.  Like before, we will examine the posterior $\beta_{1i}^c$ across each ocean to see whether or not the it is likely that, for ocean $i$, the rate of major hurricanes is increasing.  And if the posterior of $\gamma_1^c$ has most of its probability above 0, we can conclude that the rate of major hurricanes is increasing across all oceans.

The posteriors of each $\beta_{1i}^h$, $\beta_{1i}^c$ etc. are in $logit$ or {\it log odds} space, so care must be taken in direct interpretation of any estimates.  We leave them in this space so that a glace at the posterior density estimates tells the story: if most of the probabiltiy is above or below 0, we can be reasonably sure there is some effect of time.  A simple way to transform these posteriors is by taking their exponentiation: the answers then are in {\it odds} and {\it odds ratio} space.  We give some examples of this below.

\subsection{Measures of intensity}
It may be that the frequency of storms and hurricanes remains unchanged through time, but that other characteristics of these storms have changed.  One important characteristic is intensity.  We define a three-dimensional measure of intensity, in line with that defined in \citet{WebHol2005}: (1) the length $m$, in days, that a storm lives; (2) the length of the track (km) of the storm over its lifetime; and (3) the power dissipation index as derived by Emmanuel, though here we apply this to each cyclone individually and do not derive a yearly summary.

We approximate the number of days $m$ to the nearest six-hours.  Track length was estimated by computing the great circle distance between succesive six-hour observations of the cyclone, and summing these over the storm lifetime.  The power dissipation index (PDI) is defined by
\begin{equation}
\mbox{PDI} = \int_0^mV_{\max}^3dt
\end{equation}
where $V_{\max}^3$ is the maximum sustained wind speed at 10m.  Practically, we approximate the PDI---up to a constant---by summing the values $(V_{\max}/100)^3$ at each six-hour observation.  The PDI is a crude measure of the strength of the potential destructiveness of a tropical storm or hurricane, as cited by Emanuel \citeyearpar{Ema2005}.  Other than this measure, we say nothing directly about storm destructiveness (in terms of money etc.).

It was found that log transforms of these variables made them much more managable in terms of statistical analysis.  Transforming them led to all giving reasonable approximations of normal distributions; thus, standard methods are readily available.  There is substantial correlation between these three measures, which we account of in the model below.

First let, for ocean $i$, year $j$, and storm $k$ (there are $s_{ij}$ storms in year $j$ in ocean $i$),   $y_{ijk}=(\log(m)_{ijk},\log(\mbox{track length})_{ijk},\log(\mbox{PDI})_{ijk})'$, i.e. a vector quantity.  The index $l$ will denote the $l$th measure of $y$ (i.e. $y_{ijk1}=\log(m)_{ijk}$ etc).  Then we suppose that 
\begin{equation}
\log{y_{ijk}} \sim \mbox{MVN}(\mu_{ijk},\Lambda_{ij})
\end{equation}
i.e. a multivariate normal distribution where $\Lambda_{ij}$ is the $3\times 3$ precision matrix for each ocean and year. We model the mean as before
\begin{equation}
\label{eq4}
\mu_{ijkl} = \beta_{0il}^z +\beta_{1il}^z t_j + \beta_{2il}^z\mbox{CTI}_j+ \beta_{3il}^z\mbox{NAOI}_j, \:\:\: l= 1\dots3
\end{equation}
where the superscript $z$ denotes we are in the intensity portion of the model.  We further let 
\begin{equation}
\beta_{ril}^z \sim \mbox{N}(\pi_{rl},\phi_{rl}), \:\:\: r=0,1,2,3
\end{equation}
and where these hyperparameters
\begin{equation}
\pi_{rl} \sim \mbox{N}(a_{rl},b_{rl})
\end{equation}
where we use the noninformative priors $a_{rl} \sim \mbox{N}(0,1e-6), b_{rl} \sim \mbox{Gamma}(0.001,0.001)$ and 
\begin{equation}
\phi_{rl} \sim \mbox{Gamma}(0.001,0.001).
\end{equation}
As in the models for number etc., we will exmaine the marginal posterior $\beta_{1il}^z$ across each ocean and dimension $l=1,2,3$ to see whether or not the it is likely that, for ocean $i$, and dimension $l$, the mean intensity is increasing.  Also analogous to the above models, since each $\beta_{1il}^z$ has mean $\pi_{rl}$, we can examine its posterior: if it has most of its probability above 0, we can conclude that the intensity is increasing across all oceans.

Two priors were considered for the precision (inverse covariance) $\Lambda_{ij}$: a simple noninformative and a realistic, but more complicated one.  The simple (and standard) prior assumes
\begin{equation}
\label{var1}
\Lambda_{ij} \sim \mbox{Wishart}(I_{i}^3,3).
\end{equation}
i.e. a Wishart distribution with three degress of freedom and $I_{i}^3$ is the $3\times 3$ identity matrix for ocean $i$ \citep{GelCar2003}.

Below, we give evidence that there is an unambiguous, probably linear, increase in the variance of intensity through time, where the increase is proportionally equal in each of its members.  There is also substantial correlation between the parameters, which appears constant over time. The ratio of variances between members of intensity also appears constant in time.  So we built an informative prior to take these observations into account.  Let $\sigma^2_{l}$ indicate the variance of the $l$th member of intensity, and let $\rho_{lm}=\rho$ indicate the correlation, assumed constant across $l,m=1,2,3$.  Let $d_m=\sigma^2_{m}/\sigma^2_{1}$.  The variance $\sigma^2_{1}$ was modelled as
\begin{equation}
\sigma^2_{ij1} = c_0 + c_1 (t_j-1990).
\end{equation}
where $c_0$ and $c_1$ represent the intercept and slope of the observed increase in variance (subtracting 1990 centers the time and helps eliminate computer roundoff error).  Then $\sigma^2_{ij2}= d_2\sigma^2_{ij1}$ and $\sigma^2_{ij3}= d_3\sigma^2_{ij1}$.  The covariance between elements 1 and 2 was modeled as $\rho\sqrt{d_2}\sigma^2_{ij1}$: the remainder of the covariances were handled in a similar manner.  After the variance matrix $V_{ij}$ was formed, we let 
\begin{equation}
\label{var2}
\Lambda_{ij} \sim \mbox{Wishart}(V_{ij},3).
\end{equation}

In practice, with some notable exceptions mainly due to data integrity, both priors for the precision gave about equal results, except that, as expected, the second model gave results (slightly) less certain than the first.  The results were also not sensitive to exact values of $\rho,c_0$ etc. used.  They are so insensitive, in fact, that the third prior $V_{ij}=I_{ij}^3$ gave nearly identical results with (\ref{var2}): the only difference between (\ref{var2}) and (\ref{var1}) is that the former is allowed to change year by year (and over each ocean), and the later can only change over each ocean and is fixed in time.

\section{Results}
All computations were carried out in the JAGS 0.97 Gibbs sampling software \cite{JAGS} on a Fedora Core 6 platform.  The first 2000 simulations were considered ``burn in" and were removed from the analysis: 100,000 additional samples were calculated after this, with every 10th simulation used to approximate the posterior distributions (the 9 out of each 10 were discarded; this thinned the posterior simulations and helped to remove the small amount of autocorrelation of the simulations).

\subsection{Number of storms}
Figure 1 shows the number of storms $s$ from 1975-2006 for each of the five ocean basins.  There is no evident overall trend, though the number of storms seems to increase in some oceans, and decrease in others.
\begin{figure}[tb]
\includegraphics{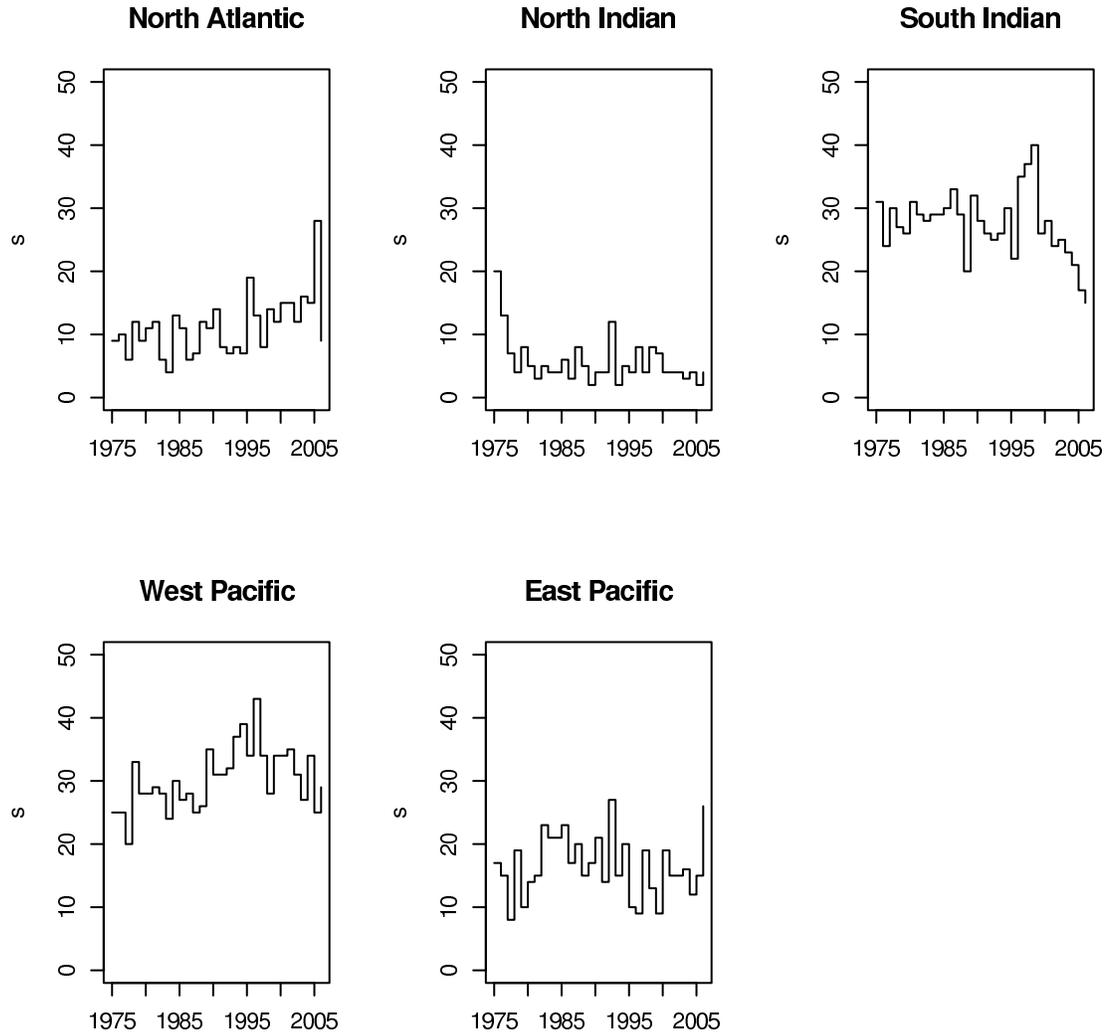}
\caption{\label{fig1} The number of storms $s$ five oceans from 1975-2006.}
\end{figure}
Other plots and other formal statistics (not shown) also indicate that the number of storms, in any given year, are independent from ocean to ocean.

The next picture shows $h/s$ for each of the oceans.  There is evidentally observational problems with the North and South Indian oceans: note the increase in the late 1970s of the ratio of hurricanes to storms.  This is almost certainly due to improvements in observations and is not entirely due to natural causes.  This being true, we nevertheless assume, for now, that the data is error free.
\begin{figure}[tb]
\includegraphics{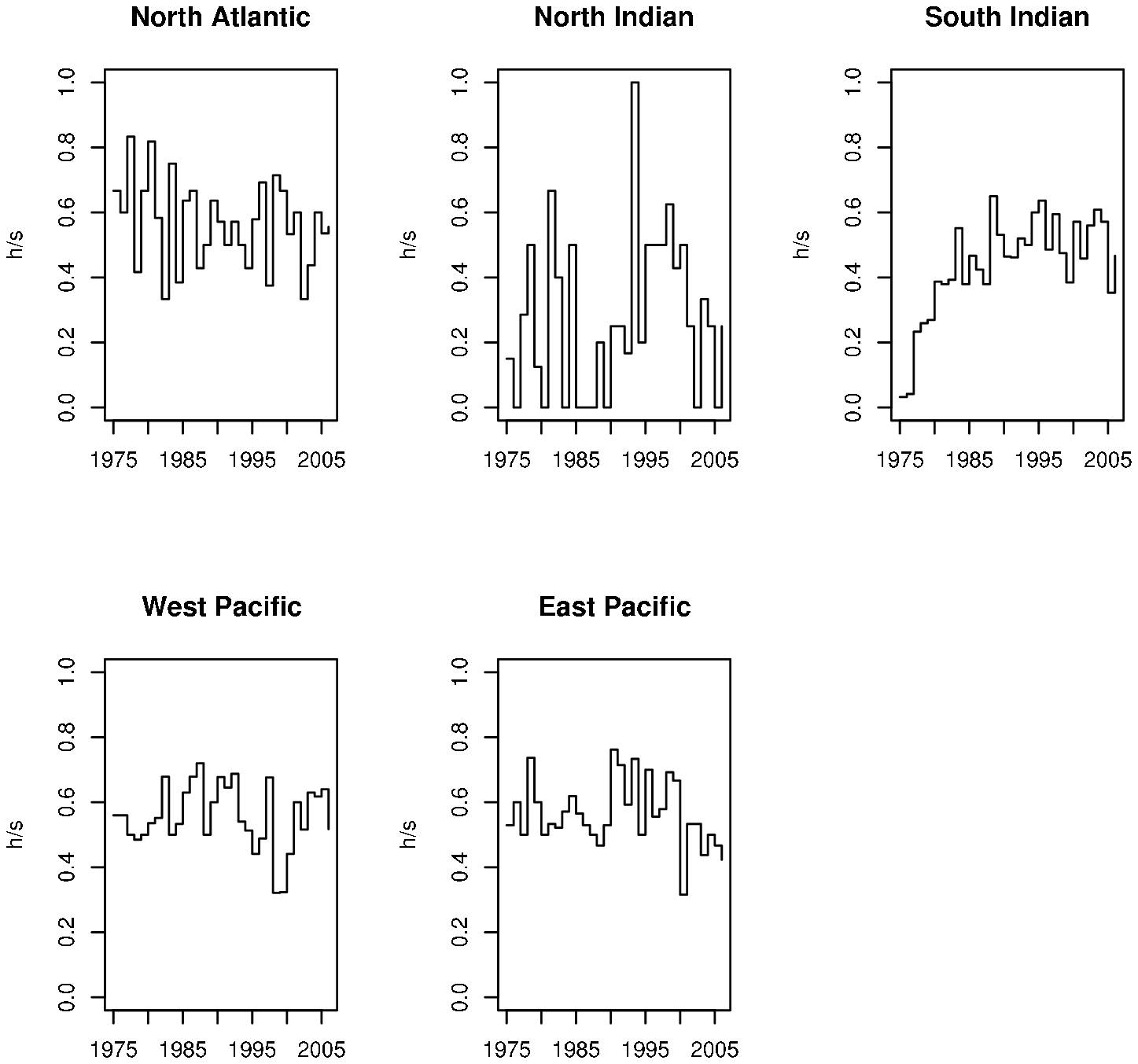}
\caption{\label{fig2} The rate $h/s$ at which tropical cyclones become hurricanes from 1975-2006.}
\end{figure}

The third picture shows $c/s$ for each of the oceans.   Again, there is no evident overall trend, and the data quality of the Indian oceans is again suspect.
\begin{figure}[tb]
\includegraphics{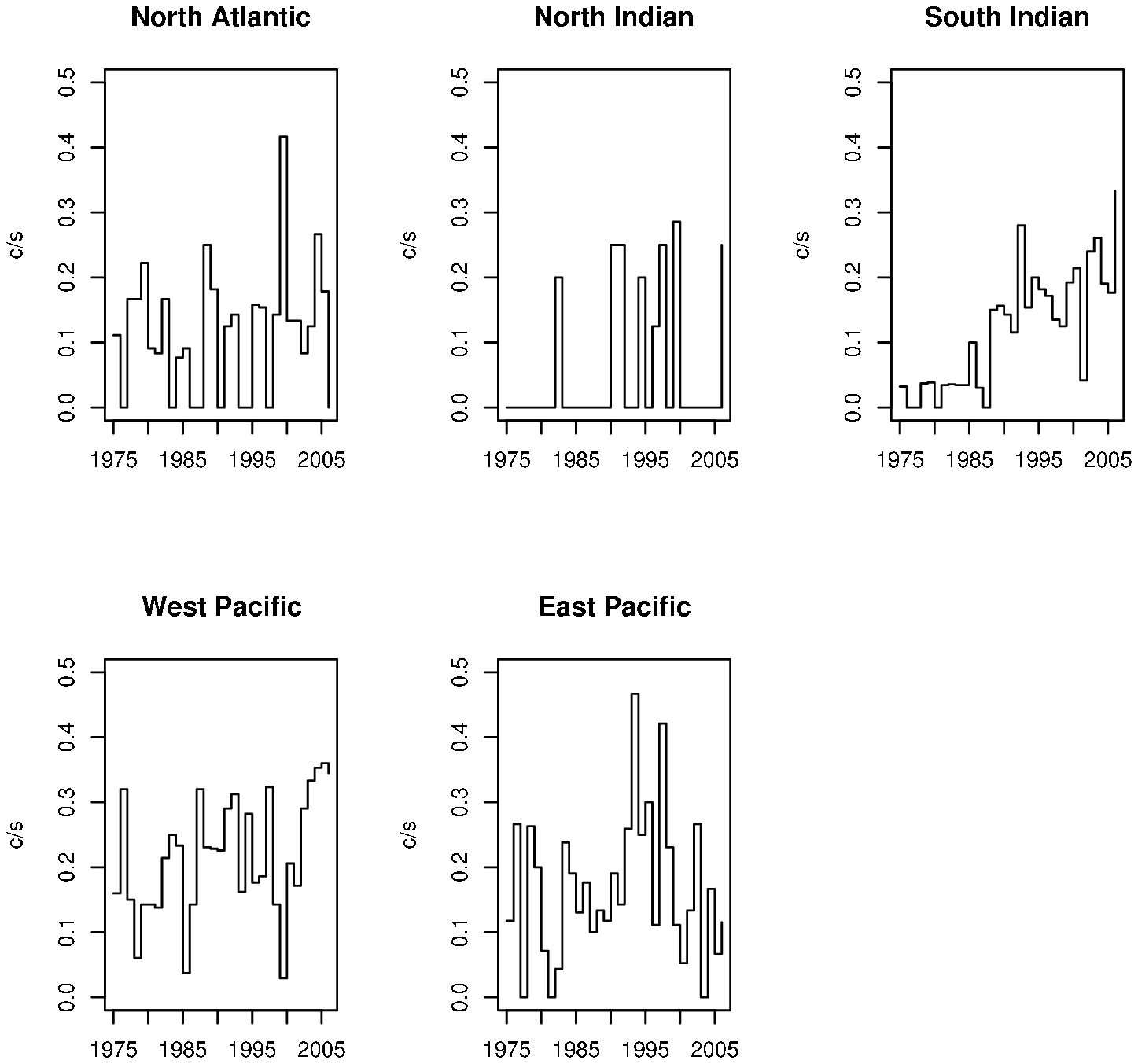}
\caption{\label{fig3} The rate $c/s$ at which tropical cyclones become category 4+ hurricanes from 1975-2006.}
\end{figure}

\begin{figure}[tb]
\includegraphics{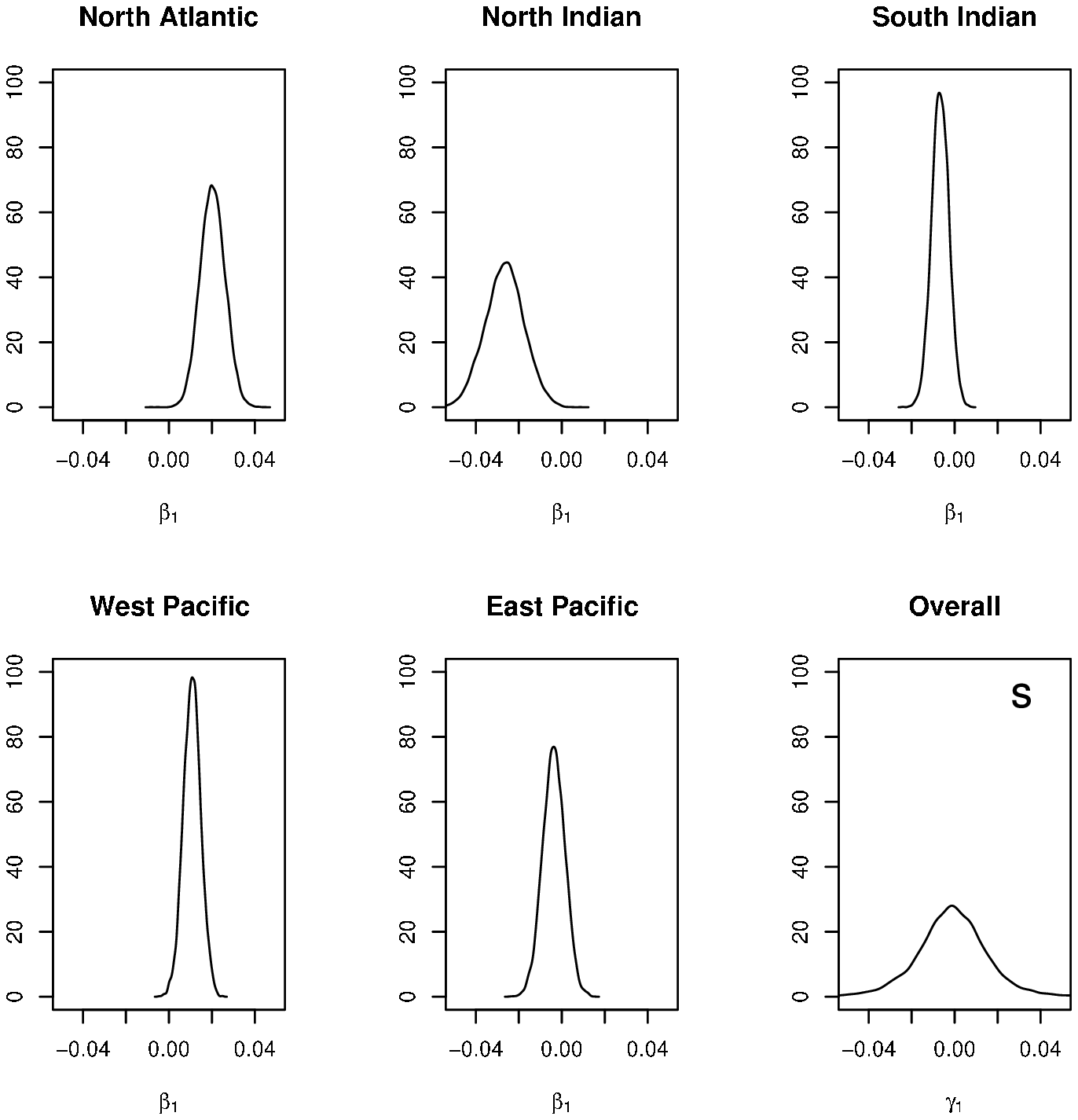}
\caption{\label{fig7} The posteriors of $\beta_{i1}$ (the regressor for change in time) and $\gamma_{1}$  (the parameter for overall change) for each ocean $i$ for the model of $s$.}
\end{figure}

\begin{figure}[tb]
\includegraphics{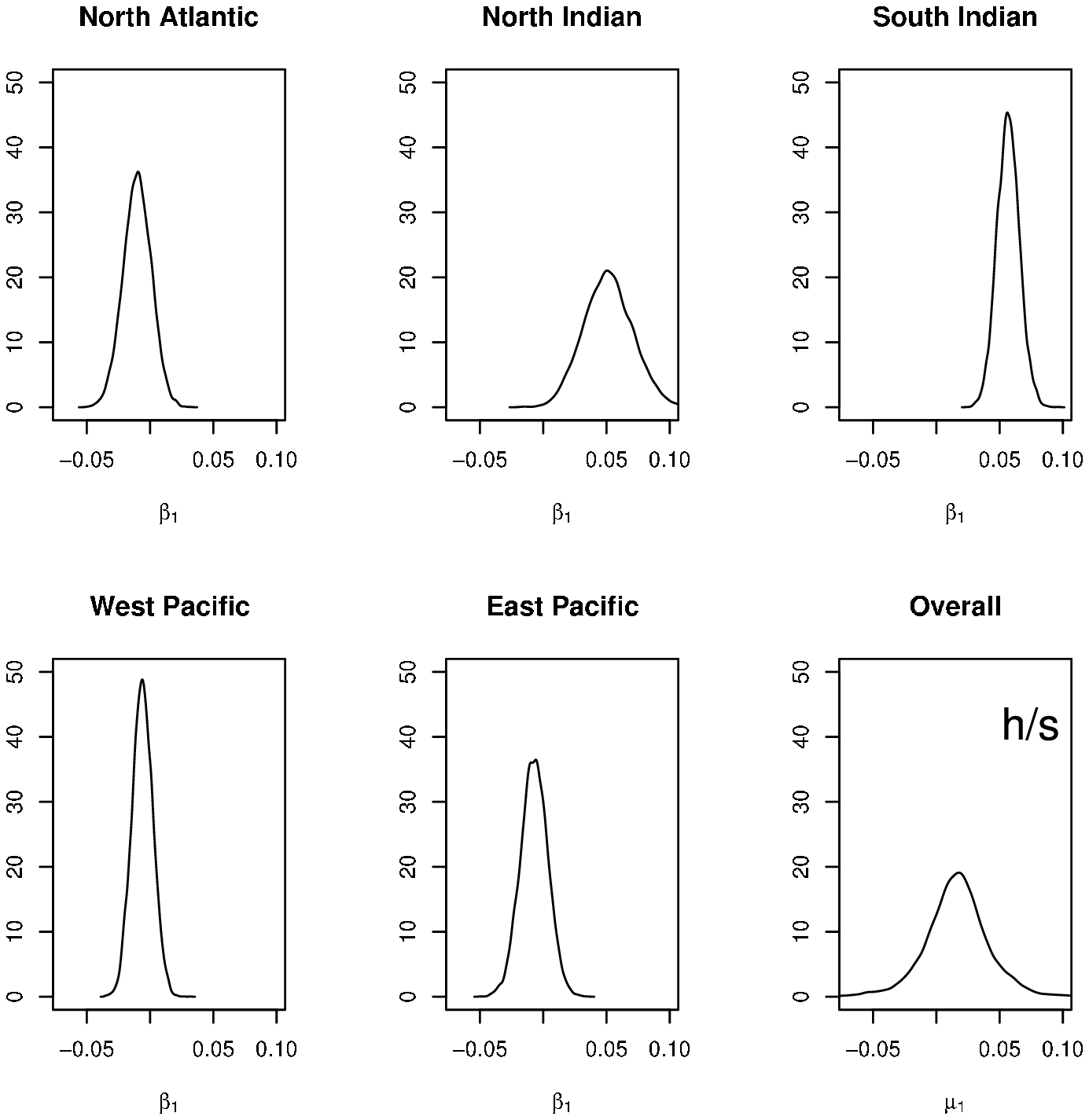}
\caption{\label{fig10} The posteriors of $\beta_{i1}$ (the regressor for change in time) for each ocean $i$ for the model of $h/s$.}
\end{figure}

\begin{figure}[tb]
\includegraphics{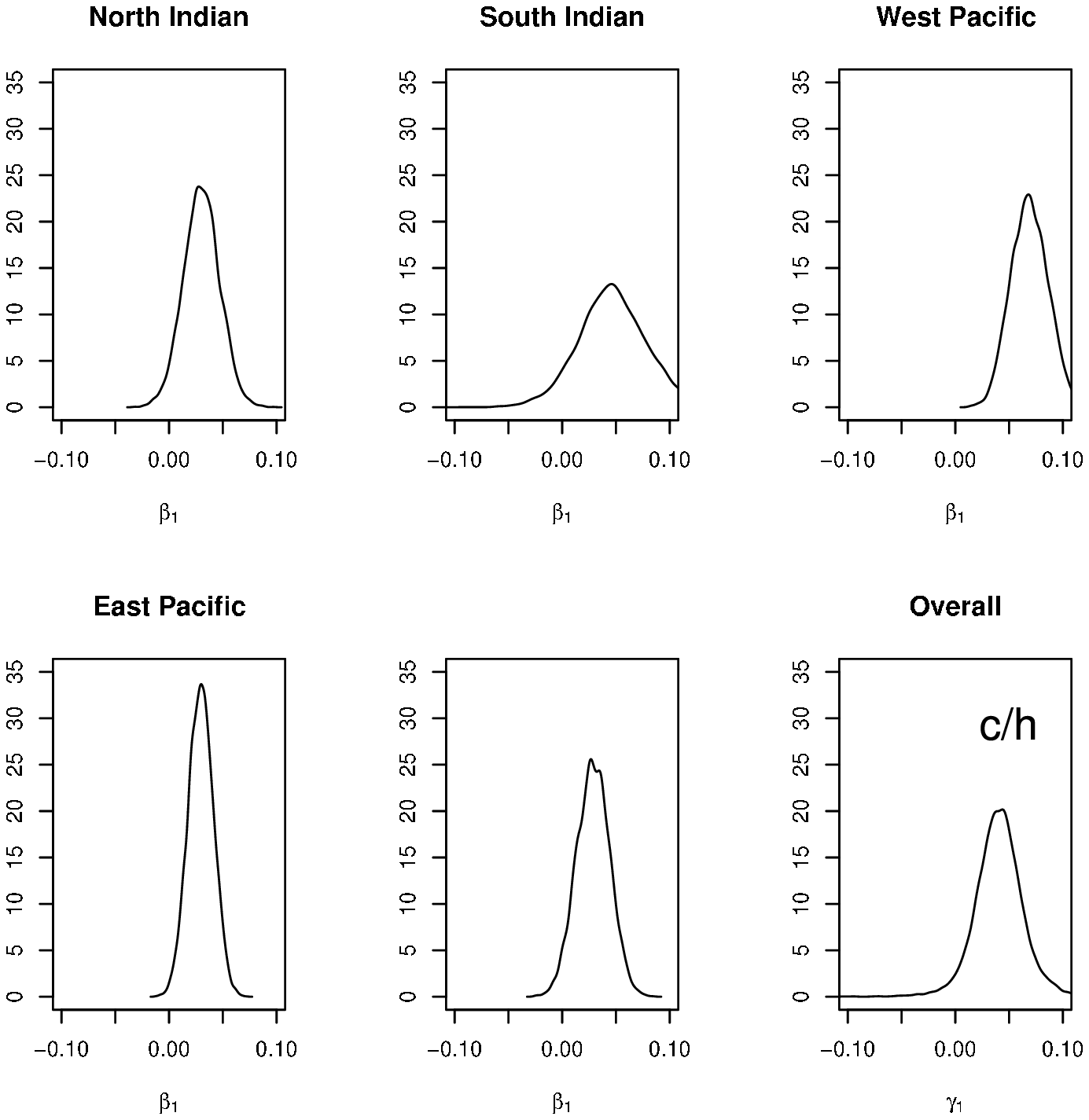}
\caption{\label{fig14} The posteriors of $\beta_{i1}$ (the regressor for change in time) for each ocean $i$ for the model of $c/h$.}
\end{figure}

The joint model for number of storms, and the ratio of evolved hurricanes and category 4+ storms was then run.  Figs. \ref{fig7}-\ref{fig14} and Tables \ref{T:beta1}-\ref{T:naoi} summarize the results.

Fig. \ref{fig7} shows there is high probability that $s$ has increased in the North Atlantic (NA) and West Pacific (WP), but decreased in the North Indian (NI), South Indian (SI), and East Pacific (EP).  Overall, as registered by the posterior of the parameter $\gamma$, there has been no overall increase or decrease.  This is backed up by the results from Table \ref{T:beta1}.  For the model of $s$, there is a greater than 97.5\% chance that there was an increase in $s$ in the NA, because all the percentiles, and in particular, the 2.5\%-tile are greater than 0.  The same is true of the WP, and the opposite is true for the NI and the SI.  Overall, there is about equal probability on either side of 0, indicating that the best guess is to say ``no change" in time.

In \citep{Bri2007} we went further and calculated the estimated probability that the posterior of  $\beta_{1}$ was greater than 0 for the NA.  It was felt that the data quality of the NA was sufficiently good enough to state the results in such precise terms.  We do not repeat these calculations here, because doing so is likely to convey more certainty than is warranted.  We repeat that the models we use are only as good as the data that go into them: the most precision we offer is in the form of the standard quantiles.

\begin{table}
\begin{center}
\caption{\it Common quantiles of the model parameter $\beta_{i1}$ and $\gamma_{1}$ (the parameters for change in time) for each ocean $i$.  Recall that the interpretion for the models $h/s$ and $c/h$ are in terms of log odds space: these numbers should be exponentiated for odds ratios. Results (increasing or decreasing) for which there is at least 95\% certainty are highlighted in bold in this and all other Tables.}
\label{T:beta1}
\begin{tabular}{lccc}
Ocean &  2.5\% & 50\% & 97.5\%  \\ \hline\hline
\multicolumn{4}{c}{$s$} \\ \hline
NA    &{\bf 0.01 }&{\bf  0.02 }&{\bf  0.03}\\
NI   &{\bf -0.04} &{\bf -0.03 }&{\bf -0.01}\\
SI   &-0.02 &-0.01 & 0.00\\
WP   &{\bf  0.00 }& {\bf 0.01} &{\bf  0.02}\\
EP   &-0.01 & 0.00 & 0.01\\
 Overall&  -0.04 &  0.00 &  0.03\\
\multicolumn{4}{c}{$h/s$} \\ \hline
NA  &-0.03 &-0.01 & 0.01\\
NI  & {\bf 0.01 } &{\bf 0.05} &{\bf  0.09}\\
SI  &{\bf  0.04 } &{\bf 0.06} &{\bf  0.07}\\
WP  &-0.02 &-0.01 & 0.01\\
EP  &-0.03 &-0.01 & 0.01\\
 Overall & -0.04 &  0.02 &  0.07\\
\multicolumn{4}{c}{$c/h$} \\ \hline
NA  &{\bf 0.00}& {\bf 0.03}& {\bf 0.06}\\
NI  &-0.02 & 0.05  &0.12\\
SI  &{\bf 0.04 }& {\bf0.07 } &{\bf 0.10}\\
WP  &{\bf 0.01 }& {\bf0.03 } &{\bf 0.05}\\
EP  &{\bf 0.00 }& {\bf0.03 } &{\bf 0.06}\\
{\bf Overall }  & {\bf 0.00}  &{\bf 0.04} &{\bf 0.09}\\
\end{tabular}
\end{center}
\end{table}

Fig. \ref{fig10}, and Table \ref{T:beta1} shows that there is some probability that the rate that hurricanes evolved from tropical storms has decreased in the NA, WP, and EP, though the result is not conculsive (there is less than a 97.5\% chance that this is true).  But there is good evidence (greater than 97.5\% chance) that the rate has increased in the NI and SI.  Whether this is due to natural causes or abberations in the data collecting mechanism, particularly in the SI, it is not possible to say.  Overall, however, the evidence is nowhere near conclusive that the rate is increasing or decreasing worldwide.

The results for the model of $c/h$ are more striking.  It is still true that the observations for the SI are problematic, but, except for the NI, there is good evidence that the rate of category 4+ storms evolving from ordinary hurricanes has increased through time in the other oceans and overall.  The odds of a category 4+ storm evolving increase, as estimated by the median overall column, by $\exp(0.04) = 1.04$ times per year: over 10 years this is an increase in the odds of 1.5 times.

\begin{table}
\begin{center}
\caption{\it The same as Table 1, but for the regressor due to the CTI.}
\label{T:cti}
\begin{tabular}{lccc}
Ocean &  2.5\% & 50\% & 97.5\%  \\ \hline\hline
\multicolumn{4}{c}{$s$} \\ \hline
NA  &{\bf -0.36 }&{\bf -0.19} &{\bf -0.04}\\
NI  &-0.27 &-0.08 & 0.08\\
SI  &-0.10 &-0.02 & 0.07\\
WP  &-0.09 & 0.00 & 0.08\\
EP  &-0.05 & 0.06 & 0.19\\
 Overall & -0.22 & -0.04 & 0.11\\
\multicolumn{4}{c}{$h/s$} \\ \hline
NA &-0.52 &-0.18 & 0.09\\
NI &-0.64 &-0.17 & 0.15\\
SI &-0.14 & 0.04 & 0.21\\
WP &{\bf 0.04 }&{\bf 0.24 }&{\bf 0.44}\\
EP &-0.23 & 0.00 & 0.21\\
 Overall  & -0.36 & -0.01 &  0.26\\
\multicolumn{4}{c}{$c/s$} \\ \hline
NA &-1.16 &-0.50 & 0.07\\
NI &-0.15 & 0.56 & 1.48\\
SI &-0.63 &-0.28 & 0.05\\
WP &{\bf 0.09} &{\bf 0.34 }&{\bf 0.59}\\
EP &{\bf 0.09} & {\bf 0.42} &{\bf 0.77}\\
 Overall  & -0.61  & 0.11 &  0.84\\
\end{tabular}
\end{center}
\end{table}

Tables \ref{T:cti} and \ref{T:naoi} show the influence that the CTI and the NAOI have on the different models.  The results here are mixed, and ocean dependent, as might be expected.  Increases in CTI are associated with a decrease the number of storms in the NA, and to a slight extent in the NI, and it tends to be associated with an increase in the number of storms in the EP.  Overall, there is not much effect.  

Similar results hold for the ratio of hurricanes to storms: increases in CTI tend to be associated with a decrease the rate at which hurricanes evolve in the NA and NI, and are associated with an increase with the rate in the WP.  We will find that the CTI, as might be expected, plays a large role for storm quality in the WP.  The SI and EP show no change; neither is there an overall effect.  Increases in CTI tend to be associated with a decrease in the rate at which category 4+ storms evolve in the NA and SI, and are strongly associated with an increase in the rate in the WP and EP.  Overall, the effect is small.

\begin{table}
\begin{center}
\caption{\it The same as Table 1, but for the regressor due to the NAOI.}
\label{T:naoi}
\begin{tabular}{lccc}
Ocean &  2.5\% & 50\% & 97.5\%  \\ \hline\hline
\multicolumn{4}{c}{$s$} \\ \hline
NA  &-0.16 &-0.04 & 0.05\\
NI  &-0.14 &-0.01 & 0.11\\
SI  &-0.13 &-0.05 & 0.03\\
WP  &-0.05 & 0.02 & 0.10\\
EP  &-0.06 & 0.03 & 0.14\\
 Overall & -0.11 & -0.01 &  0.08\\
\multicolumn{4}{c}{$h/s$} \\ \hline
NA &-0.31 &-0.08 & 0.08\\
NI &-0.46 &-0.10 & 0.07\\
SI &-0.14 &-0.01 & 0.15\\
WP &-0.16 &-0.02 & 0.12\\
EP &-0.24 &-0.06 & 0.09\\
 Overall  & -0.24 & -0.06 &  0.08\\
\multicolumn{4}{c}{$c/h$} \\ \hline
NA &{\bf   0.04} &{\bf   0.26} &{\bf   0.53}\\
NI &-0.02 & 0.25 &  0.59\\
SI &{\bf   0.06 }&{\bf  0.26} &{\bf  0.47}\\
WP &{\bf   0.00 }&{\bf 0.21 }& {\bf 0.40}\\
WP &{\bf   0.03 }&{\bf   0.25} &{\bf   0.48}\\
{\bf Overall } &{\bf  0.05} &{\bf  0.25} &{\bf  0.46}\\
\end{tabular}
\end{center}
\end{table}

The NAOI has little association on either the number of storms or the rate at which these evolve into hurricanes.  But it does have a stronger association on rate at which category 4+ storms evolve: in every ocean, and overall, the effect is positive: an increase in a higher NAOI is  associated with an increase in the rate at which strong hurricanes evolve.

\subsection{Measures of intensity}
There was one missing track length (out of 1756) in the South Indian ocean. There were 28 (out of 622) missing PDIs in the North Indian and 106 (out of 1756) in the South Indian. The multivariate model we used requires that be no missing values.  So we imputed the small number of missing values by forming linear regression models of log track (or log PDI) as a function of log $m$ and log PDI (or log $m$ and log track), and used these to make predictions at the missing values.  We do not expect that this unduly effected any of the results.

\begin{figure}[tb]
\includegraphics{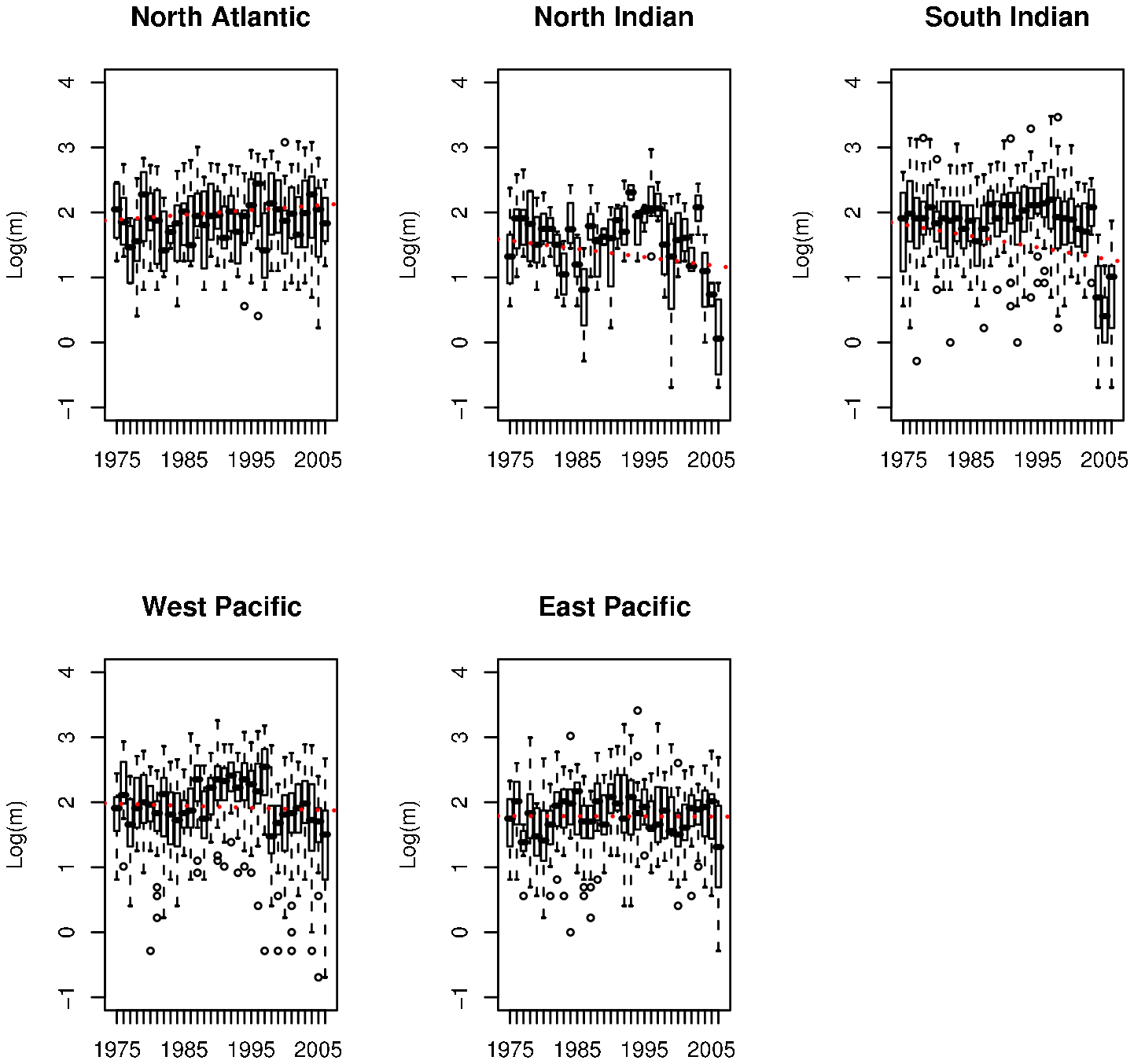}
\caption{\label{fig4} The time series of boxplots, for each year and each ocean, of $\log(m)$.  A simple dashed trend line of the medians in overlayed.}
\end{figure}

\begin{figure}[tb]
\includegraphics{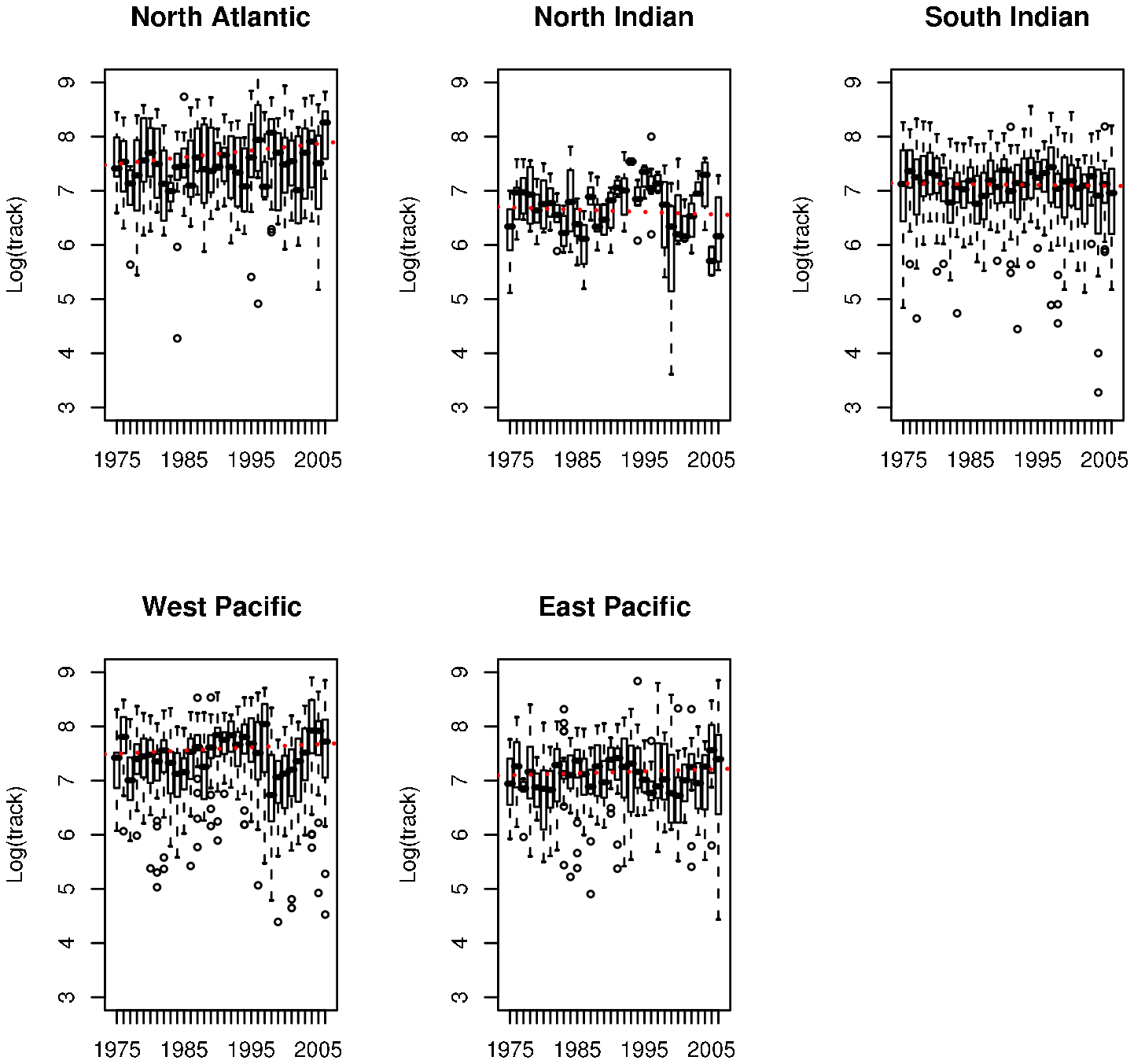}
\caption{\label{fig5} The time series of boxplots, for each year and each ocean, of $\log(\mbox{track length})$.  A simple dashed trend line of the medians in overlayed.}
\end{figure}

\begin{figure}[tb]
\includegraphics{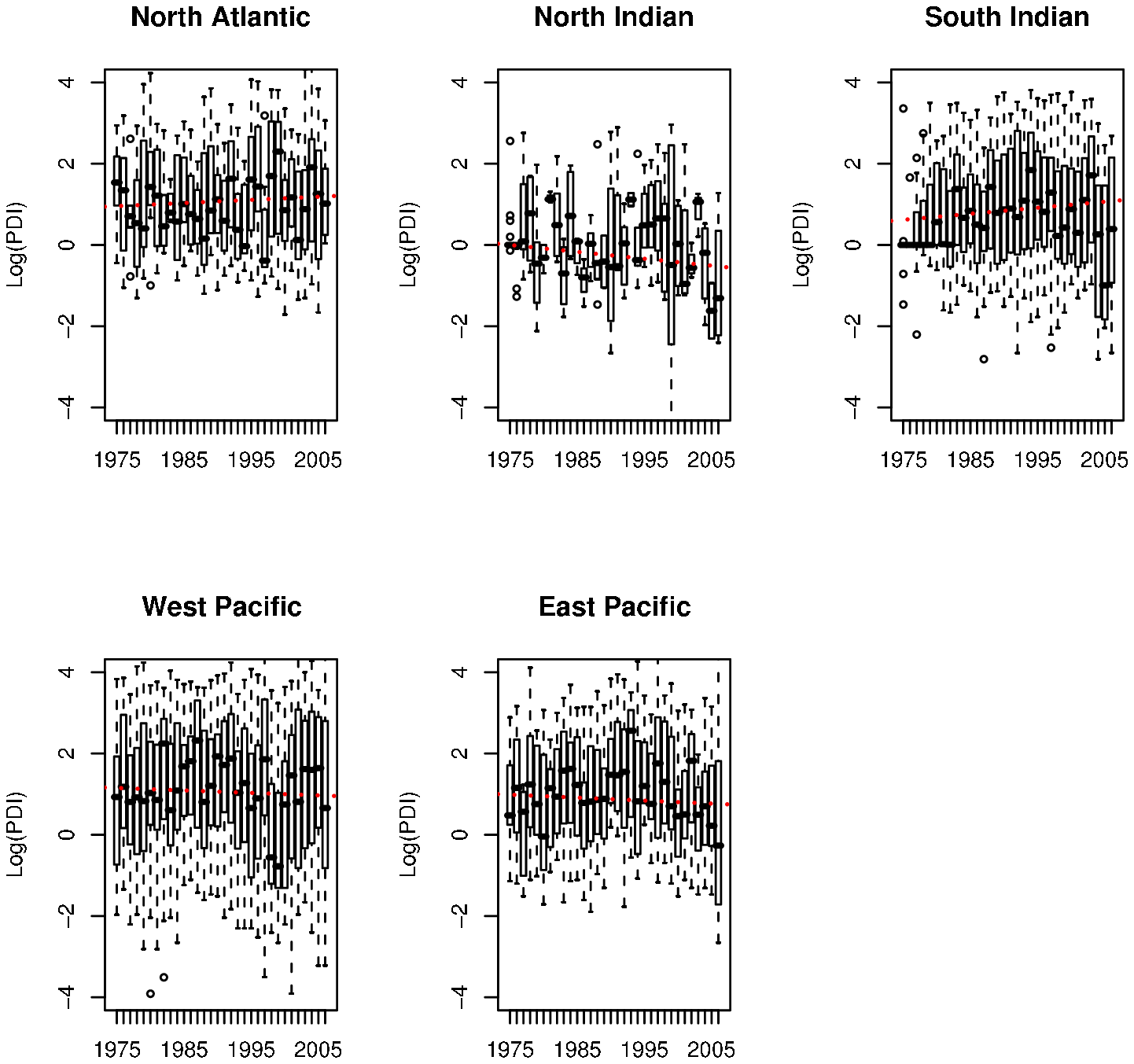}
\caption{\label{fig6} The time series of boxplots, for each year and each ocean, of $\log(\mbox{PDI})$.  A simple dashed trend line of the medians in overlayed.}
\end{figure}

Figures \ref{fig4}-\ref{fig6} show the boxplot time series of individual storm (within a year and logged) $m$, track length, and PDI for each ocean.  A simple regression trend line for the medians of these distributions is overlayed to help guide the eye as to the overall trend, if any. There is still the same data quality problem with the Indian oceans, but there are some trends evident in these pictures.

\begin{figure}[tb]
\includegraphics{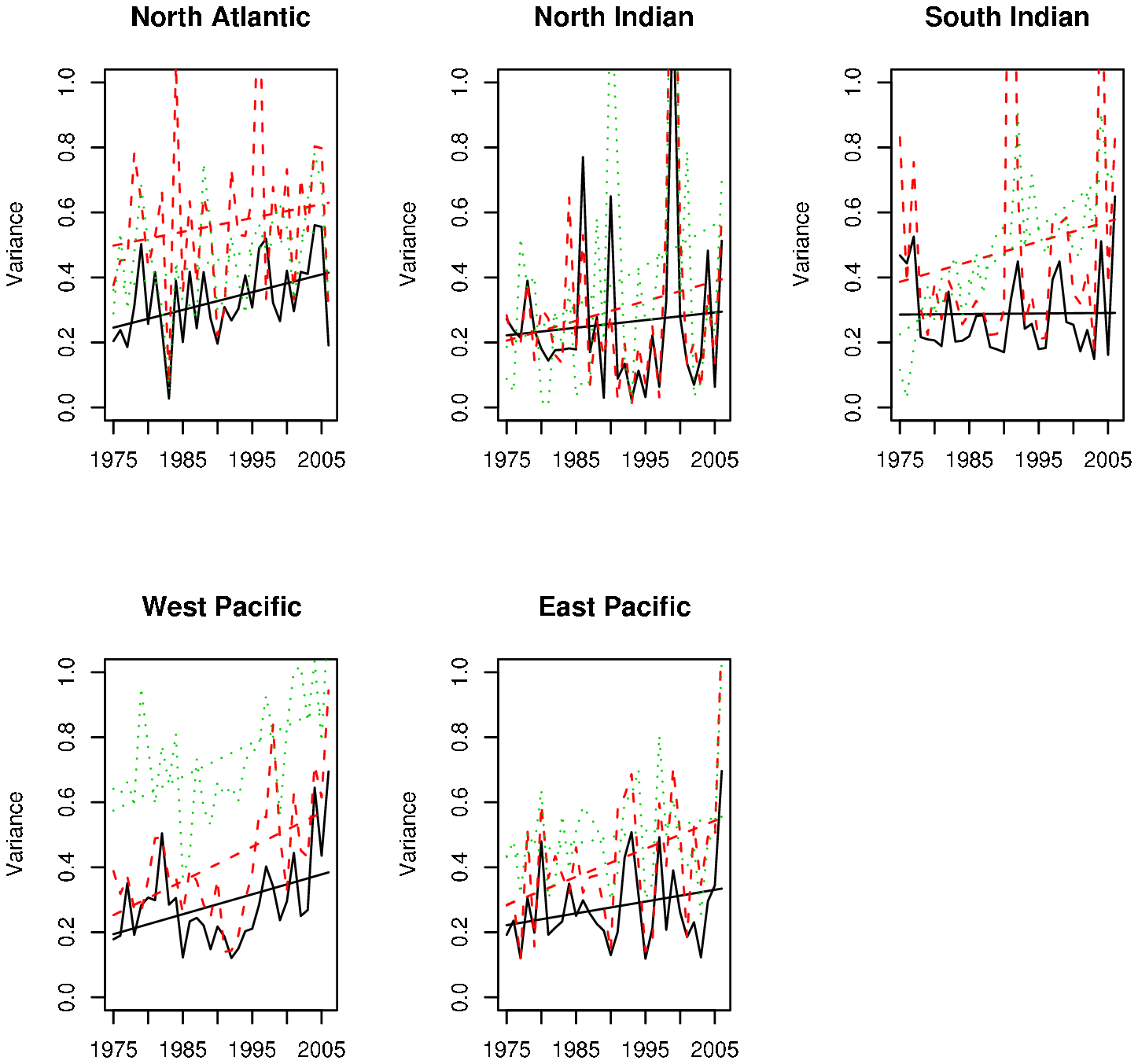}
\caption{\label{fig13} The estimated variance of $\log(m)$ (solid line), log(track) (dashed), and log(PDI) (dotted).  Log(PDI) has been scaled by dividing by 4 so that all data fits on one picture.  The is a clear increase in variability of each measure.}
\end{figure}

Within each year we estimated the covariance between intensity and plotted, in Fig. \ref{fig13}, the estimated variance of $\log(m)$ (solid line), log(track) (dashed), and log(PDI) (dotted).  Log(PDI) has been scaled by dividing by 4 so that all data fits on one picture.  The is a clear increase in variance of each measure in Fig. \ref{fig13}, similar across all oceans.  Certainly, there is a lot of noise around this trend, particular (again) in the Indian oceans, but its existence is unambiguous.  This represents one of our main findings.

\begin{figure}[tb]
\includegraphics{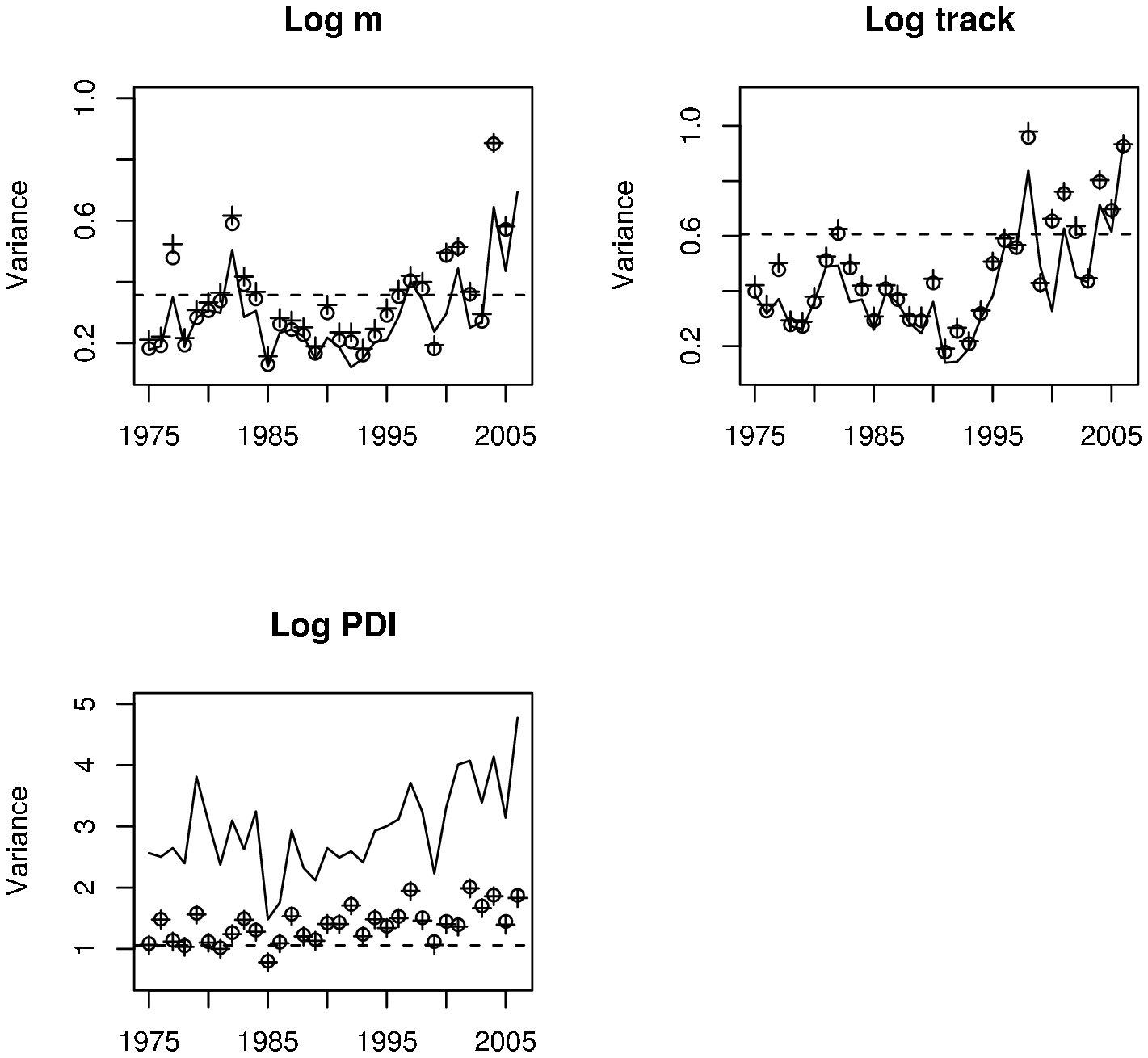}
\caption{\label{fig20} For the Western Pacific, four estimates of the year-by-year variance of intensity: the classical point estimate (solid line); posterior median based on the increasing variance  prior (\ref{var2}) (open circles); the posterior median based on the non-informative year-by-year prior; the posterior median based on the simple ocean prior (\ref{var2}) (dashed line).}
\end{figure}

To illustrate this in a different way, we present Fig. \ref{fig20}, which is four estimates of the year-by-year variance of intensity for the Western Pacific: the classical point estimate (solid line); posterior median based on the increasing variance  prior (\ref{var2}) (open circles); the posterior median based on the non-informative year-by-year prior; the posterior median based on the simple ocean prior (\ref{var2}) (dashed line).  The posterior medians differ trivially for the priors (\ref{var2}) and the simplified year-by-year version.  The prior (\ref{var1}) does not track very well with the other priors in this ocean, but it does better in other oceans, particularly the Indian (not shown).  The Bayesian and classical estimates are in very close agreement for $\log(m)$ and $\log(\mbox{track})$, but differ markedly for $\log(\mbox{PDI})$: this is also found in the other oceans: the difference appears largest in this ocean, the ocean which also appeared to have the clearest increase in variance over time.  This exhibits the typical Bayesian ``shrinkage" that is found in multivariate estimation.  Note: the classical (solid line) should certainly {\it not} be taken as the true value of the variance to which the Bayesian estimate aspires.  It is merely another estimate, one that may be too high (just as the Bayesian estimate may be too low).  In any case, all methods show a clear increase in variance through time, which is all that we are claiming holds.

From the data in Fig. \ref{fig13}, we estimated the parameters of the second precision prior (\ref{var2}).  We estimate (by averaging the classical estimates over the oceans): $\widehat\rho=0.8\:\: (.11)$, $\widehat{d}_2=2\:\: (.56)$, $\widehat{d}_3=6\:\: (1.7)$, and $\widehat{c}_0=0.3\:\: (0.03)$, $\widehat{c}_1=0.004\:\: (0.0026)$.  The numbers in parentheses are the estimated standard deviations.  The model results presented below are very insensitive to the exact values of these parameters.  It turns out to be much more important whether the variance is allowed to change through time by each ocean, or that it is the same through time at each ocean.  That is, we repeated all the analyses below, but this time with $V_{ij} = I_{ij}$, i.e. nearly the same as the non-informative `flat' prior $V_{ij} = I_{i}$, except we allow the variance to change year-by-year: the results below barely changed when this different, but eminently fair, prior was used.

\begin{figure}[tb]
\includegraphics{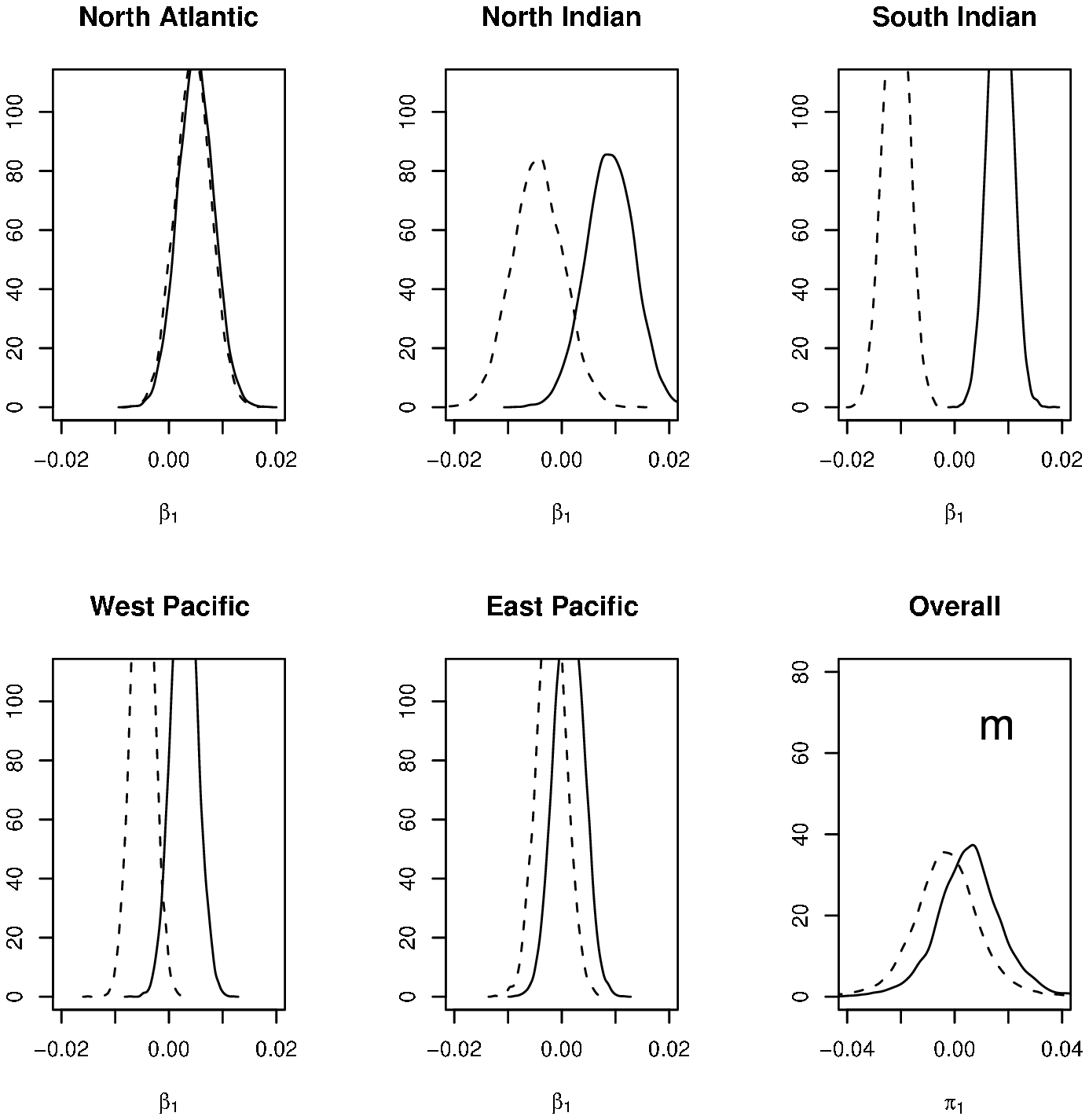}
\caption{\label{fig17} The marginal posteriors of $\beta_1$ and $\pi_1$ for $\log(m)$ across all oceans.  The solid line represents the posteriors with the increasing variance prior; the dashed line is for the simple, constant non-informative prior.  As expected, in each ocean, the increasing variance prior produced wider---that is, less certain---posteriors.  The results for both priors, however, are in rough agreement; but see the text for a discussion of the South Indian ocean.}
\end{figure}

Figure \ref{fig17}, and Table \ref{T:intensity}, shows the posteriors of $\beta_{1i1}^z$ and $\pi_{r1}$ for $\log(m)$ for each variance prior.  The posterior based on the increasing variance prior (\ref{var2}) is the solid line; the posterior based on the ``flat" or constant non-informative prior (\ref{var1}) is the dashed line. There is no difference in the NA: both priors indicate that there may be some evidence that $\log(m)$ has increased, but we cannot be very sure this is so.  The same conclusion can be reached for the EP.  No change is seen in the West and East Pacific.

But for the North and South Indian, a complete opposite picture emerges (also partly seen in the West Pacific), particularly for the South Indian (where the data is most suspect).  The posterior based on the flat prior (\ref{var1}) gives good evidence that $\log(m)$ has decreased, but the posterior based on the increasing prior (\ref{var2}) gives good evidence that $\log(m)$ has increased!  This is partly because of the strong correlation between $\log(m)$ and $\log(\mbox{PDI})$, and the strong rise over time of the later: this is picked up in prior (\ref{var2}).  We emphasize that this is not because of some bias in the informative increasing prior: a nearly identical picture arises if we instead use the year-by-year flat prior.

Now, the raw data picture of $\log(m)$ in the Indian oceans show decreases, but there is a strong rise in the $\log(\mbox{PDI})$ in the South Indian, which has about 3 times as many data points as does the North Indian ocean (and would thus receive more weight in the simulations).  These factors go a long way into explaining the difference in interpretation.  In any case, any result for the Indian Oceans should be viewed with caution because these results are so sensitive on the prior used.

\begin{figure}[tb]
\includegraphics{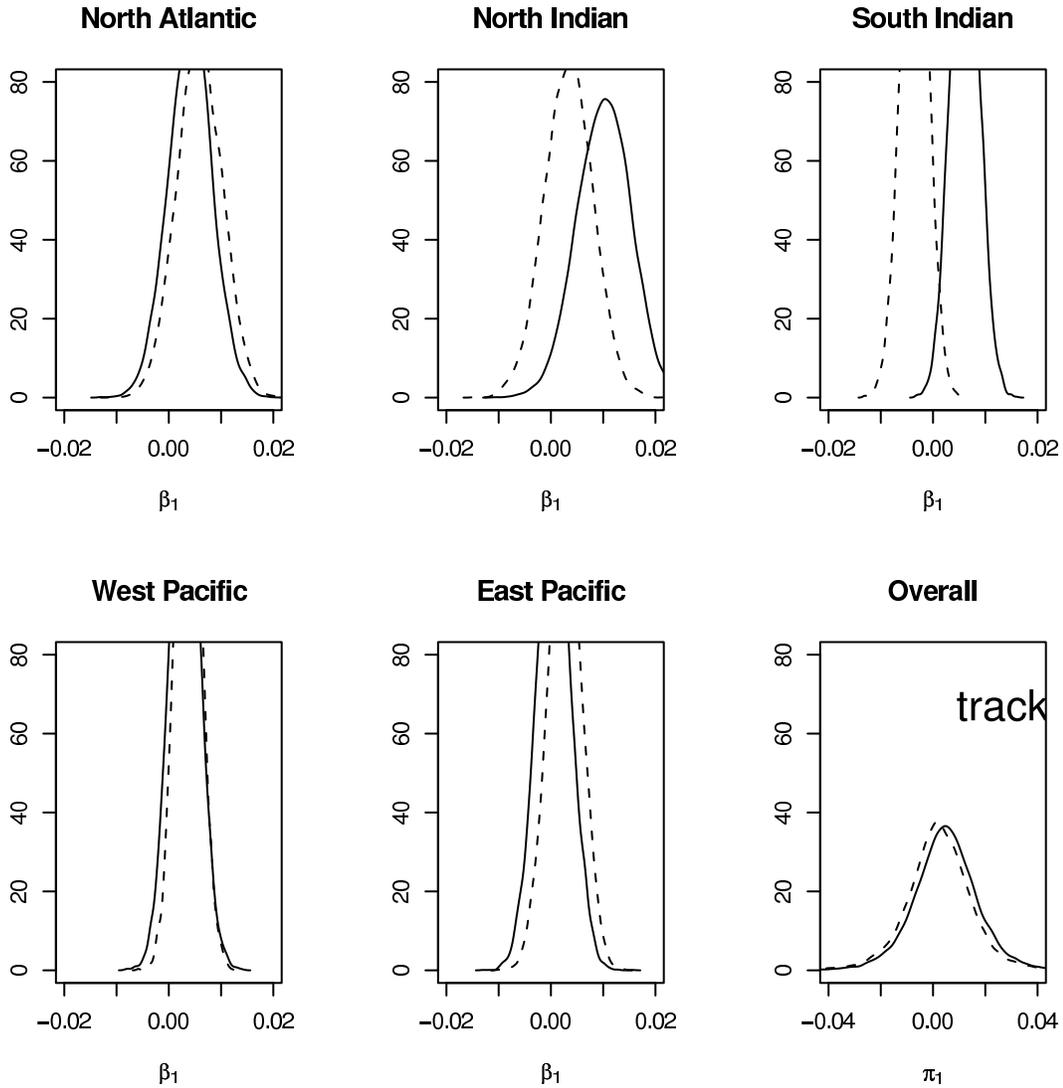}
\caption{\label{fig18} The same as Fig. \ref{fig17} but for $\log(\mbox{track})$.  Here, both variance priors give closer agreement.}
\end{figure}

The overall trend parameter for $\log(m)$ is near 0, meaning it is likley that $\log(m)$ has neither increased nor decreased across all oceans.

The same interpreation for $\log(\mbox{track})$ and $\log(\mbox{PDI})$ in Figs. \ref{fig17} and \ref{fig18} as for $\log(m)$ can be made.  The two variance priors give nearly identical results, except for the Indian Oceans; and there is no overall trend.

\begin{table}
\begin{center}
\caption{\it Common quantiles of the model parameter $\beta_{i1}$ and $\gamma_{1}$ (the regressor for change in time) for each ocean $i$.  Results are for both the increasing variance and non-informative (`flat') priors.  There do not appear to be any overall trends.}
\label{T:intensity}
\begin{tabular}{lcccccc}
      & \multicolumn{3}{c}{Increasing $V$} & \multicolumn{3}{c}{Flat $V$}\\ \hline
Ocean &  2.5\% & 50\% & 97.5\%  & 2.5\% & 50\% & 97.5\% \\ \hline\hline
\multicolumn{7}{c}{$\log(m)$} \\ \hline
NA &-0.001  &0.005 &0.012	& -0.002 & 0.004 & 0.011\\
NI &{\bf 0.000  }&{\bf 0.009} &{\bf 0.018}& -0.014 & -0.005 & 0.005\\
SI &{\bf 0.004 } &{\bf 0.008} &{\bf 0.013}	&{\bf  -0.016} &{\bf -0.011 }&{\bf -0.006}\\
WP &-0.002 &0.003 &0.008	&{\bf  -0.009} &{\bf -0.005} &{\bf -0.001}\\
EP &-0.004 &0.001 &0.006	& -0.007 &-0.002 & 0.003\\
 Overall & -0.020 &  0.005 & 0.031 & -0.030 & -0.003& 0.023\\
\multicolumn{7}{c}{$\log(\mbox{track})$} \\ \hline
NA &-0.004 &0.004 &0.012	 &-0.003&  0.006& 0.014\\
NI &{\bf 0.000 }&{\bf 0.010} &{\bf 0.024} &-0.006&  0.003& 0.012\\
SI &{\bf 0.001 }&{\bf 0.006 }&{\bf 0.020} &-0.009& -0.004& 0.001\\
WP &-0.003 &0.003 &0.009	 &-0.001&  0.004& 0.008\\
EP &-0.006 &0.001 &0.007	 &-0.004&  0.003& 0.009\\
 Overall & -0.021 & 0.005 & 0.031 & -0.024 &  0.002&  0.028\\
\multicolumn{7}{c}{$\log(\mbox{PDI})$} \\ \hline
NA &-0.004  &0.007 &0.019	 &-0.006 & 0.005 &0.016\\
NI &-0.002  &0.010 &0.022	 &-0.008 & 0.002 &0.012\\
SI & {\bf 0.019 } &{\bf 0.027} &{\bf 0.034 }&{\bf  0.007 }&{\bf  0.014 }&{\bf 0.021}\\
WP &-0.006  &0.004 &0.014	 &-0.007 & 0.001 &0.009\\
EP &-0.011 &-0.001 &0.009	 &-0.013 &-0.003 &0.006\\
 Overall & -0.019 & 0.009 & 0.037 & -0.023 &  0.004&  0.031\\
\end{tabular}
\end{center}
\end{table}

\begin{figure}[tb]
\includegraphics{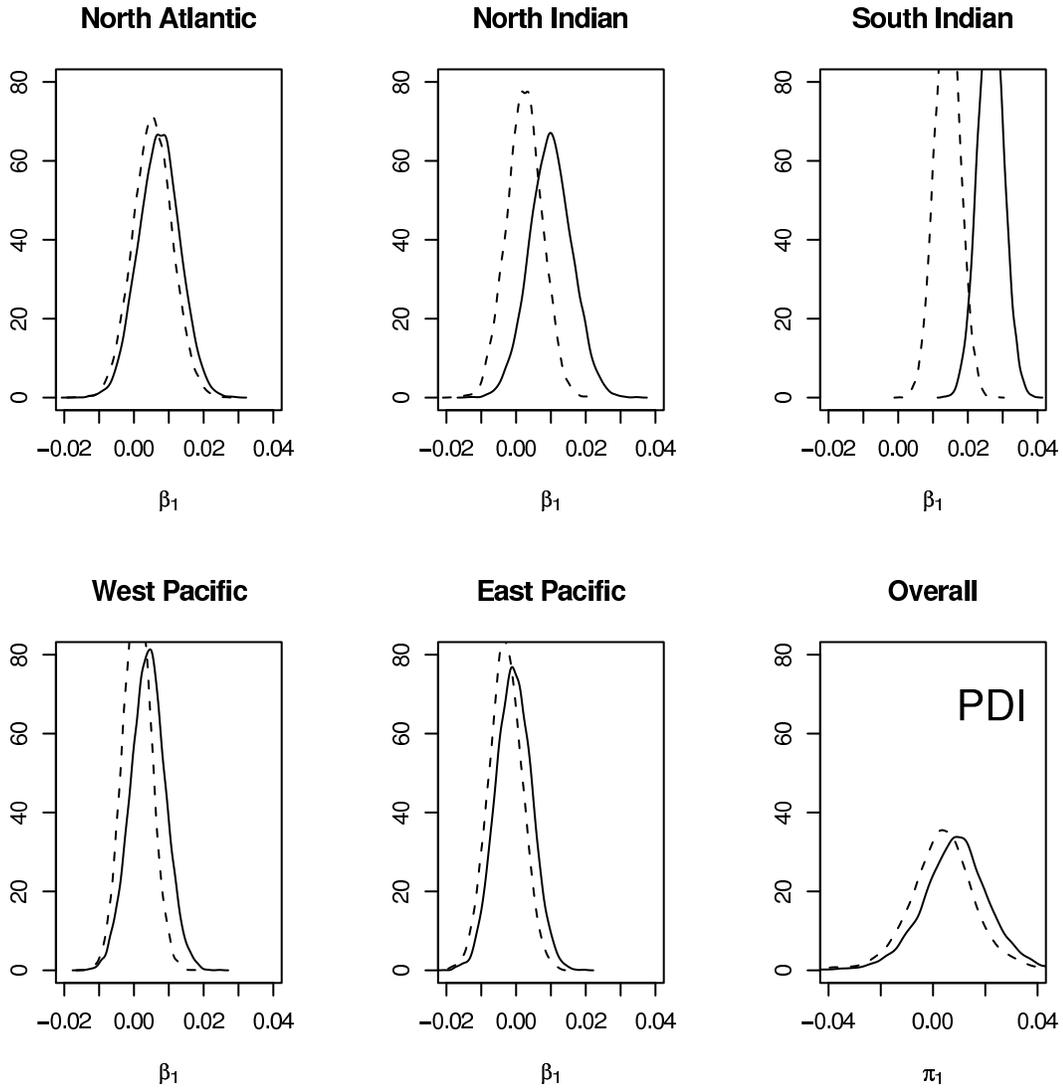}
\caption{\label{fig19}  The same as Fig. \ref{fig17} but for $\log(\mbox{PDI})$.  Here, both variance priors give closer agreement except for the South Indian ocean.}
\end{figure}

\begin{table}
\begin{center}
\caption{\it Common quantiles of the model parameter $\beta_{i2}$ and $\gamma_{2}$ (the regressor for influence of CTI) for each ocean $i$.  Results are for both the increasing variance and non-informative (`flat') priors.}
\label{T:intensityCTI}
\begin{tabular}{lcccccc}
      & \multicolumn{3}{c}{Increasing $V$} & \multicolumn{3}{c}{Flat $V$}\\ \hline
Ocean &  2.5\% & 50\% & 97.5\%  & 2.5\% & 50\% & 97.5\% \\ \hline\hline
\multicolumn{7}{c}{$\log(m)$} \\ \hline
NA &-0.11 &-0.02 & 0.06	 &-0.10 &-0.01 & 0.07\\
NI &-0.01 & 0.08 & 0.16	 &-0.03 & 0.06 & 0.16\\
SI &-0.05 & 0.00 & 0.05	 &-0.03 & 0.03 & 0.08\\
WP &{\bf 0.14} &{\bf 0.19} &{\bf 0.25}&{\bf 0.09 }&{\bf 0.14} &{\bf 0.19}\\
EP &-0.02 & 0.04 & 0.10	 &-0.03 & 0.03 & 0.08\\
 Overall & -0.04 &  0.06 & 0.15 & -0.03 & 0.05& 0.13\\
\multicolumn{7}{c}{$\log(\mbox{track})$} \\ \hline
NA &-0.10 & 0.00 & 0.10	&-0.08 &0.03 & 0.13\\
NI &{\bf  0.05 }& {\bf 0.14} &{\bf  0.23}& {\bf 0.07}&{\bf 0.16}& {\bf  0.27}\\
SI &-0.08 &-0.02 & 0.04	& -0.05& 0.01&  0.07\\
WP &{\bf  0.15} & {\bf 0.22 }&{\bf  0.29}	& {\bf  0.12}&{\bf  0.19}&{\bf  0.25}\\
EP &{\bf 0.00 }& {\bf 0.08} &{\bf  0.15}&  {\bf 0.00}& {\bf 0.07}& {\bf  0.14}\\
 Overall & -0.04 & 0.07 & 0.18 & -0.02 &  0.08&  0.17\\
\multicolumn{7}{c}{$\log(\mbox{PDI})$} \\ \hline
NA &-0.28 &-0.12 & 0.03	 &-0.27 &-0.11 & 0.03\\
NI &-0.21 &-0.09 & 0.04	 &-0.19 &-0.06 & 0.05\\
SI &-0.09 &-0.01 & 0.07	 &-0.05 & 0.03 & 0.11\\
WP &{\bf  0.14} &{\bf  0.26 }&{\bf  0.38	} &{\bf  0.11 }& {\bf 0.22} &{\bf  0.33}\\
EP &-0.05 & 0.06  &0.18	 &-0.02 & 0.09 & 0.20\\
 Overall & -0.10 & 0.04 & 0.18 & -0.07 &  0.05&  0.17\\
\end{tabular}
\end{center}
\end{table}

Table \ref{T:intensityCTI} shows how the CTI is related to intensity.  For each dimension of intensity, a higher CTI is associated with greater intensity in the WP, and to a smaller extent in the EP.  To be clear: a high CTI means longer lived, longer travelling, and windier storms in the Western Pacific.  Longer travelling storms in the NI are also associated with higher CTI.

\begin{table}
\begin{center}
\caption{\it Common quantiles of the model parameter $\beta_{i3}$ and $\gamma_{3}$ (the regressor for influence of NAOI) for each ocean $i$.  Results are for both the increasing variance and non-informative (`flat') priors.}
\label{T:intensityNAOI}
\begin{tabular}{lcccccc}
      & \multicolumn{3}{c}{Increasing $V$} & \multicolumn{3}{c}{Flat $V$}\\ \hline
Ocean &  2.5\% & 50\% & 97.5\%  & 2.5\% & 50\% & 97.5\% \\ \hline\hline
\multicolumn{7}{c}{$\log(m)$} \\ \hline
NA &-0.07 &-0.02 & 0.03	& -0.05 &0.00 & 0.05\\
NI &-0.03 & 0.03 & 0.09	& -0.04 &0.02 & 0.09\\
SI &-0.06 &-0.02 & 0.02	& -0.03 &0.01 & 0.05\\
WP & {\bf 0.01 }& {\bf 0.05 }&{\bf  0.09	}& {\bf  0.00 }&{\bf 0.04 }& {\bf 0.08}\\
EP &-0.06 &-0.01 & 0.03	& -0.04 &0.00 & 0.04\\
 Overall & -0.05 &  0.00 & 0.06 & -0.03 & 0.01& 0.06\\
\multicolumn{7}{c}{$\log(\mbox{track})$} \\ \hline
NA &-0.10 &-0.02 & 0.04	 &-0.08 &-0.01 & 0.05\\
NI &-0.05 & 0.02 & 0.09	 &-0.06 & 0.00 & 0.06\\
SP &-0.04 & 0.01 & 0.05	 &-0.03 & 0.01 & 0.06\\
WP &-0.02 & 0.03 & 0.09	 &-0.02 & 0.02 & 0.06\\
EP &-0.03 & 0.02 & 0.07	 &-0.03 & 0.02 & 0.07\\
 Overall & -0.05 & 0.01 & 0.06 & -0.04 &  0.01&  0.06\\
\multicolumn{7}{c}{$\log(\mbox{PDI})$} \\ \hline
NA &-0.07&  0.01 & 0.11	& -0.08&  0.00 & 0.09\\
NI &-0.15& -0.04 & 0.04	& -0.11& -0.02 & 0.05\\
SI &-0.04&  0.03 & 0.10	& -0.03&  0.03 & 0.10\\
WP &-0.10& -0.01 & 0.06	& -0.07&  0.00 & 0.07\\
EP &-0.09&  0.00 & 0.08	& -0.07&  0.00 & 0.07\\
 Overall & -0.07 & 0.00 & 0.06 & -0.06 &  0.00&  0.07\\
\end{tabular}
\end{center}
\end{table}

The association of intensity with NAOI is presented in Table \ref{T:intensityNAOI}.  There is little association in any ocean or overall, with the, perhaps surprising, exception that higher NAOI is associated with longer lived storms in the Western Pacific.

\section{Conclusions}

We find that there is good evidence that the number of tropical cyclones over all the oceans basins considered here have neither increased nor decreased in the past thirty years: some oceans saw increases, others decreases or no changes.  These results stands even after controlling for CTI and NAOI.  These results are of course conditional on the model we used being adequate or at least it being a reasonable approximation to the data.  As in our first paper, we make no predictions about future increases as it would be foolish to extrapolate the simple linear models we used into the future.  

We also found that the rate at which tropical cyclones become hurricanes does {\it not} appear to be changing through time across oceans, nor is it much influenced by CTI or NAOI. The rate may be increasing in the Indian oceans, but it seems just as likely that flaws in the data would account for the results we have seen. 

There is good evidence that the mean rate at which major (category 4 or above) storms evolve from ordinary hurricanes has increased through time, though the increase is small.

We find almost no evidence that the mean of the distribution of individual storm intensity, measured by storm days, track length, or individual storm PDI, has changed (increased or decreased) through time over all the oceans.  Again, there were certain noted increases in the Indian oceans, which may be real or may be due to flaws in the data: this is evidenced by the posteriors from these oceans being very sensitive to the priors used.  We did, however, find an unambiguous increase in the variance of the distribution of storm intensity over all oceans.

It might be asked, if the overall number of storms has stayed constant, but the rate of category 4+ storms has increased, which implies that the number of category 4+ storms has increased, why has not, say, the mean of PDI (which is a direct function of wind speed) increased because of these stronger storms?   This is most likely because the variance of PDI {\it has} increased, meaning there has been an increase of both stronger {\it and} weaker storms, but that this change has balanced (in the sense that the mean has stayed the same).  Of course, the variance of $m$ and track length have also increased, leading to both shorter and longer lived, and shorter and longer travelling storms.

The CTI was particularly associated with storm quality in the Western Pacific: the rate at which hurricanes and category 4+ storms evolve increased with higher CTI.  Higher NAOI also led, worldwide, to an increase in the rate at which category 4+ storms evolve.  Higher CTI was also associated with longer and more powerful storms in the Western, and to a smaller extent, the Eastern Pacific oceans.  Perhaps surprising, was that higher NAOI was associated with longer lived storms in the Western Pacific and nowhere else.

In the Introduction we noted the hurricane reanalysis project of Kossin et al. \citeyearpar{KosKna2007}.  The methods used here could certainly be used in this, and other datasets like it.  These analysis from them would make a valuable comparison to the results presented here.

\bibliography{ams}

\end{document}